# A colossal impact enriched Mars' mantle with noble metals

**R. Brasser[1†] and S. J. Mojzsis[2,3†]**

[1]Earth Life Science Institute, Tokyo Institute of Technology, Ookayama, Meguro-ku, Tokyo 152-8550, Japan.

[2]Department of Geological Sciences, University of Colorado, UCB 399, 2200 Colorado Avenue, Boulder, Colorado 80309-0399, USA

[3]Institute for Geological and Geochemical Research, Research Centre for Astronomy and Earth Sciences, Hungarian Academy of Sciences, 45 Budaörsi Street, H-1112 Budapest, Hungary.

†Collaborative for Research in Origins (CRiO), The John Templeton Foundation – FfAME Origins Program

Corresponding author: Stephen J. Mojzsis (mojzsis@colorado.edu)

**Key Points:**

- Mars experienced a large impact in its first ca. 130 Myr.
- This event is most likely the primary source of the highly siderophile element abundance in the martian mantle.
- Mars' crustal hemispherical dichotomy and satellites may be a visible remnant of this impact.

**Abstract**

Once the terrestrial planets had mostly completed their assembly, bombardment continued by planetesimals left-over from accretion. Highly siderophile element (HSE) abundances in Mars' mantle imply its late accretion supplement was 0.8 wt.%; Earth and the Moon obtained an additional 0.7 wt.% and 0.02 wt.%, respectively. The disproportionately high Earth/Moon accretion ratio is explicable by stochastic addition of a few remaining Ceres-sized bodies that preferentially targeted Earth. Here we show that Mars' late accretion budget also requires a colossal impact, a plausible visible remnant of which is the hemispheric dichotomy. The addition of sufficient HSEs to the martian mantle entails an impactor of at least 1200 km in diameter to have struck Mars before ca. 4430 Ma, by which time crust formation was well underway. Thus, the dichotomy could be one of the oldest geophysical features of the martian crust. Ejected debris could be the source material for its satellites.

**1 Introduction**

Terrestrial planet formation, which includes such details as the effects of late accretion on the final compositional makeup of the planets, remains a long-standing problem in planetary science. In traditional dynamical models the terrestrial planets grow from a coagulation of planetesimals into protoplanets [Kokubo & Ida, 1998]. These objects eventually become dynamically unstable and subsequently evolve into a phase wherein planetary mergers through large impacts (or sometimes 'giant impacts' in the sense of Earth's Moon-forming event) were commonplace. Consequently, these protoplanetary collisions eventually led to the observed terrestrial planet inventory [e.g. Chambers & Wetherill, 1998]. One of the predicted outcomes of giant impact models is that the Moon, Earth and other inner solar system planets ought to have experienced a long history of what is termed 'late accretion'. That is, after core formation and initial separation of the silicate reservoirs (crust/mantle), left-over planetesimals on planet-crossing orbits are consumed by the terrestrial bodies as substantial mass supplements [Wetherill, 1975]. These events strongly modulated the preservation of silicate crusts on the terrestrial planets, and as such also provide a robust temporal bounds on the earliest sustainable appearance of a biosphere in the inner solar system.

Evidence for such addition is inferred from the abundance of highly siderophile elements (HSEs; Re, Os, Ir, Ru, Pt, Rh, Pd, Au) in the mantles of Earth, Moon and Mars. The proportions are chondritic relative to each other but greatly exceed their expected abundance after metal-silicate partitioning [Kimura et al., 1974]. This indicates that either HSEs were delivered after core formation and silicate differentiation [Chou, 1978], or were left behind by inefficient metal–silicate partitioning at higher pressures and temperatures [e.g. Murthy, 1991, Righter et al., 2015]. Here we explore the implications of the former.

The relative abundance accreted by the planets should scale as the ratio of their gravitational cross sections, if it is assumed that all the planetesimals have the same size and are far more numerous than the planets. For the Earth and the Moon this ratio is approximately 20. Instead, HSE abundance in the terrestrial and lunar mantles suggest that Earth experienced approximately two orders of magnitude more late accretion than the Moon, contrary to that proposed by theory [Day et al., 2007; Day & Walker, 2015].

The high ratio of the terrestrial and lunar HSE mantle budgets led Bottke et al. [2010] to conclude that the size-frequency distribution of the remaining planetesimals from planet formation had to have been shallow even at large diameters (D≥500 km), and the majority of the mass delivered to the Earth came from a few large objects comparable in size to the asteroid 1 Ceres (diameter ~950 km) that yet lurked within the inner solar system. The low population number of these large objects leads to a stochastic impact regime that statistically favors collisions with Earth [Sleep et al., 1989].

The late accretion model to explain Earth's HSE budget is a subject of on-going debate [e.g. Touboul et al., 2012, 2014; Willbold et al., 2011, 2015; cf. Righter et al., 2015]; this dispute is no-less vigorous for purported late accretion to Mars. Early analyses of HSEs in martian meteorites suggested a late addition of 0.1 to 1 wt.% to Mars [Kong et al., 1999; Warren et al., 1999]. Walker [2009] reported that Mars experienced a mass supplement of ~0.7 wt.%, which is comparable to the 0.3 to 0.8 wt.% proposed for Earth [Day et al., 2016]. Interpretations of osmium isotope values in martian meteorites seem to back this assertion by indicating that Mars accreted at least 0.5 wt.% of material of chondritic composition after core formation [Brandon et al., 2012; Day & Walker, 2015], with the most likely contribution being approximately 0.8 wt.% [Day et al., 2016].

A recent analysis of metal-silicate partitioning for the platinum group elements (HSEs minus Au and Re) in martian meteorites, combined with theoretical partitioning models used to construct inverse models of Mars' mantle composition, shows that the concentration of these rare metals in the martian mantle cannot be solely established by metal-silicate equilibration early in the planet's history [Righter et al., 2015] because the long-term time-integrated Os concentration would be sub-chondritic, in violation of the evidence from martian meteorites. In short, the amount of late accretion on Mars is still questioned, but given its existence on the Moon and the Earth and what we understand about the nature of late accretion and the dynamics in the inner solar system, it makes sense that Mars experienced something similar.

Recently Brasser et al. [2016] combined *N*-body and Monte Carlo simulations to establish the nature of late accretion onto the terrestrial planets, and the likely total mass accreted by the Earth and Mars when calibrated to lunar values. That work could not reproduce the high ratio between the mass accreted by the Earth versus the Moon at greater than 99% confidence if the diameter of the largest planetesimals did not exceed 2000 km. The implication is that Earth's late accretion came predominantly from a single impact with a lunar-sized (~3500 km diameter) object between ca. 4500 Ma and 4400 Ma which delivered most of the HSEs presently lodged in the terrestrial mantle. In an alternative view, Sleep [2016] suggested that the HSEs were delivered during the giant impact (GI) that formed the Moon, obviating the need for a late accretion component. This assertion is in contradiction with dynamical models [e.g. Wetherill 1975; Raymond et al., 2013] and geochemical analyses. The analysis of Brasser et al. [2016] neglected to survey the possible nature of late accretion on Mars, however, and to assess the likelihood that a single impactor could explain the martian HSE abundances. In this study, we investigate that scenario in more detail.

Here, we combined *N*-body and Monte Carlo simulations from Brasser et al. [2016] with the results of new Monte Carlo simulations to explain Mars' HSE budget. In our discussion of implications of our models for a singular large impact, we explicitly set out predictions of our results in the context of martian composition and dynamical regimes, including the origin of the crustal dichotomy and the martian moons, Phobos and Deimos.

**2 Methods**

The Monte Carlo impact simulations undertaken for this study consisted of the following: Planetesimals with diameters that range from 1 km to 2000 km were randomly generated. The size-frequency distribution of the planetesimals was defined to be identical to that of the main asteroid belt because the oldest (>4.4 Ga) surfaces on the Moon and Mars, and possibly Mercury, exhibit the same shape of their distributions [Strom et al., 2005; Fasset et al., 2012; Werner, 2014]. This is a top-heavy distribution and the majority of the mass resides in the largest bodies. It was assumed that each planetesimal has a density of 3000 kg m$^{-3}$ and its mass is added to the total mass in planetesimals.

The total mass added to the terrestrial planets and the Moon depends on the amount of mass in remnant planetesimals from terrestrial planet formation, and on the impact probability of the planetesimals with the planets. From N-body simulations of the inner solar system after the Moon-forming impact Brasser et al. [2016] compute the average impact probability of a planetesimal is 12% with Earth, 1.9% with Mars and 0.5% with the Moon, respectively. The lunar value is barely affected by tidal evolution owing to the fact that the Moon's gravitational focusing is negligible. If in the Monte Carlo simulation a planetesimal strikes one of these three target bodies, its mass is added to the total accreted mass because for simplicity we assume perfect accretion. This assumption agrees with previous *N*-body simulations that included imperfect accretion [Raymond et al., 2013]. Once this is accomplished, the next planetesimal is produced in our Monte Carlo simulation. In these time-independent simulations, Brasser et al. [2016] continued to generate planetesimals until the Moon had accreted at least 0.025 wt.%, corresponding to estimates of its late accretion complement [Day et al., 2007; Day & Walker, 2015].

Mars, however, is demonstrably older than the Moon [Dauphas & Pourmand, 2011; Tang & Dauphas, 2014] so that it correspondingly had more time to accrete its mantle HSE budget. For this reason additional Monte Carlo simulations were performed as part of this project. The simulations were terminated once Mars reached a threshold where at least 0.5 wt.% was added to account for the calculated accretion component [Day & Walker, 2015].

## 3 Results

Before we proceed to investigate late accretion to Mars we quickly review an important result from [Brasser et al., 2016]. **Figure 1** shows the diameter of the largest planetesimal to strike the Moon (grey), Mars (red) or the Earth (blue) versus the total accreted mass reported in weight % of the target's mass. Closer examination of the lunar data shows that the largest planetesimal accreted by the Moon in our previous model has a diameter of approximately 200 km, consistent with size of the impactor required to account for the diameter of the South Pole-Aitken basin [Potter et al., 2012]. Data show that even when Earth was struck with at least one "planetesimal" with a diameter of 2000 km, the total amount of mass accreted by Earth in the late veneer is at best 0.4 wt.%. Extrapolation of the Earth data indicates that a "planetesimal" diameter of about 3000 km is required to have the Earth accrete another 0.7 wt.% subsequent to the Moon-forming event. This extrapolation led Brasser et al. [2016] to conclude that the late veneer on the Earth was most likely the result of a single, large impact with a lunar-sized (~3000 km diameter) body.

For Mars, the outcome is somewhat more complicated compared to the Earth-Moon case. If we begin with the assumption that all of Mars' late accretion was delivered *after* the Earth's Moon had already formed, then Mars should have experienced a collision with a planetesimal whose diameter exceeded 1000 km.

The results of the new Monte Carlo simulations, where Mars had to have accreted a minimum of 0.5 wt.%, are depicted in **Figure 2**. Here, we show that the minimum diameter of the largest object to have struck Mars is about 500 km if it experienced only 0.5 wt.% late accretion, whereas this object had to have been at least 1200 km in diameter to yield 0.8 wt.% late accretion. A roughly Ceres- (or somewhat larger) sized planetesimal is expected have been mostly accreted by Mars, with only a small fraction of the mass escaping [Raymond et al., 2013]. The impact of a planetesimal with a diameter of 1200 km will deliver approximately 0.5 wt.% of material to Mars. For the shallow size-frequency distribution that was employed in our work, approximately a third of the total delivered mass is in the largest planetesimal [Tremaine & Dones, 1993]. The rest of the mass (in this case 0.3 wt.%) is delivered by smaller bodies.

Even though crater scaling laws are usually not applicable to basins, we can still use the number of known martian basins to implicitly constrain the size of the largest impactor. All that is required is the diameter of the smallest planetesimal that will excavate a basin on Mars; any planetesimals that strike Mars larger than this size will automatically form additional basins. There are at least 30 large impact basins that can be confidently documented on Mars [e.g. Schultz & Frey, 1999; Werner, 2014]. In **Figure 3** we plot the number of expected basins on Mars versus the size of the largest impactor from our Monte Carlo numerical simulations. If we can only account for approximately 30 basins, **Figure 3** and the simulation data show that the largest impactor in the field of possibilities in the late accretion Monte Carlo simulations and tied to the HSE abundance has a diameter of approximately 1300 km.

In summary, if Mars did experience an approximately 0.8 wt.% increase of its mass by late accretion, then this must account for (at least) one large body with a diameter of at minimum 1200 km. Importantly, our model outcome coincides with the largest expected impactor described in Marinova et al. [2008] and Bottke et al. [2010]. Strangely enough such an outcome cannot account for Mars' approximately 25 h rotation period . The expected number of sidereal rotations of a planet per revolution around the Sun following an episode of stochastic accretion is $\eta=(R_H/R_M)^{3/2}(m/M_M)$ [Dones & Tremaine, 1993], where $R_M$ is Mars' radius, $R_H$ is Mars' Hill radius, m is the mass of the impactor and $M_M$ is Mars' mass. Currently $R_H/R_M \sim 450$ and an impactor with D=1200 km and Mars' mean density has $m/M_M \sim 0.003$ so that $\eta \sim 30$. Thus, the impact that contaminated Mars' mantle is expected to have only contributed about 5% of Mars' spin angular momentum.

**4 Discussion**

We now explore whether the amount of late accretion on Mars computed here is consistent with the expected mass in planetesimals from leftover terrestrial planet formation [cf. Quintana et al., 2016].

Data strongly suggest that Mars is a stranded planetary embryo which formed in less than 10 Myr [Dauphas & Pourmand, 2011; Tang & Dauphas, 2014]. Mars formed in a region with a relatively low density of planetesimals [Brasser et al., 2017] whereas Earth and Venus were still busy accreting planetesimals and planetary embryos. Brasser et al. [2016] determined that at the time of Moon-formation, the mass in remnant planetesimals was of the order of a fifth of a lunar mass (or ca. 0.002 Earth masses). The Monte Carlo simulations performed here keep track of the total number and mass of planetesimals (see Methods). For Mars to have accreted an additional 0.8 wt.% we compute that there was on average 0.04 Earth mass in planetesimals on Mars-crossing orbits in the inner solar system (otherwise the planetesimals cannot impact Mars) by using the typical impact probability of a planetesimal with Mars. In other words, the remaining planetesimal mass is computed as $m_p=m_{imp}/P_M$, where $m_{imp}$ is the mass that impacted Mars (0.8 wt.%) and $P_M$ is the probability of impacting Mars (about 2%). Mars is thought to have had a magma ocean phase that lasted approximately 100 Myr [Debaille et al., 2007; cf. Humayun et al., 2013]; in our view it is easier, though not necessary, to incorporate the HSEs into the martian mantle before solid crust formation. We further suggest the low mass in planetesimals occurred by about 4.55 Ga, by which time Mars had nearly fully formed [Dauphas & Pourmand, 2011; Tang & Dauphas, 2014]. This mass value is consistent with a few times $10^{-3}$ Earth masses in planetesimals remaining at the time of the Moon-forming event when accounting for dynamical ejection and collisions with the terrestrial planets [Brasser et al., 2016].

If Mars indeed experienced a large late veneer-scale impact early in its history, what evidence could possibly exist for such an event in its lithosphere? Unlike the Earth or Venus, Mars has undergone relatively little resurfacing; the southern half of its surface is characterized by an ancient (~4.3 – 4.4 Ga) cratered terrane [Werner et al., 2014] while the northern hemisphere shows evidence of some form of later crustal processing and extensive volcanism, and its corresponding surface age is only 3.7 Ga [Werner et al., 2011]. These points lead us to conclude that Mars probably avoided the sort of planet-wide devastating impact that melted the crust and mantle down to great depth in conditions comparable in magnitude (equi-mass impactor:target) to that which created Earth's Moon.

Others have proposed that Mars' hemispheric dichotomy came about from a giant impact, the most obvious expression being the northern lowlands [Wilhelms & Squyres, 1984; Andrews-Hanna et al., 2008; Nimmo et al., 2008; Marinova et al., 2008]. Based on geometric arguments, the impactor to account for this purported structure had a diameter of about 1000 km, if the collision occurred head-on [Wilhelms & Squyres, 1984], or 1600-2700 km in diameter for an oblique collision that comports relatively well with the dichotomy's geometry and ellipticity [Nimmo et al., 2008; Marinova et al., 2008]. Our results outlined in the previous section are in agreement with these various size estimates for a giant impact, and because it appears necessary for Mars to have experienced a large impact to be consistent with its mantle HSE budget, it is well within the realm of possibility that the martian hemispherical dichotomy is the result of this giant impact.

The timing of this event, however, is uncertain. Wilhelms & Squyres [1984] suggest that the colossal impact that created the hemispherical dichotomy occurred before 4.4 Ga. Nimmo et al. [2008] preferred a similar timing because of the suggested longevity of the martian magma ocean [Debaille et al., 2007; cf. Humayun et al., 2013]. On the other hand, Harper et al. [1995] were the first to report an excess of $^{142}$Nd in martian meteorites and used these data to conclude that differentiation of the silicate reservoirs – the last time that the crust and mantle of Mars were equilibrated from complete melting – occurred before 4.54 Ga. They suggest that Mars experienced no giant impacts afterwards. Foley et al. [2005] reached the same conclusion from coupled $^{142}$Nd and $^{182}$W anomalies in martian meteorites and Borg et al. [2016] uses neodynium isotopes to derive an age of 4.5 Ga for the formation age of the martian basaltic meteorites. It is difficult to imagine, however, that an impact this early in Mars' history would leave a lasting imprint on its crust. The oldest known martian zircon is found in NWA 7533, a sibling of NWA 7034 [Agee et al., 2013], with an age of 4.428 Ga [Humayun et al., 2013]. This age coincides with the end of the purported long-lived martian magma ocean, and if it is interpreted to be a minimum age of crust formation we conclude that the impact that created the hemispherical dichotomy occurred before about 4.43 Ga.

Mars is a small, old and cold world that ineffectively recycles its crust in a stagnant-lid regime over geologic timescales [Yin et al., 2012 and references therein]. A mass augmentation event of the magnitude we propose to the crust and mantle of primordial Mars from a chemically and isotopically distinct source ought to impart a geochemical signal in the form of strong isotopic heterogeneities in different ancient martian silicate reservoirs. Over billion-year timescales, we further expect that these implanted isotopic heterogeneities would become gradually subdued or homogenized with bulk Mars by slow crustal processing such as intra-crustal partial melting and sluggish convection. Earlier, we insinuated that a very early (ca. 4.5 Ga) impact would most likely result in a homogenization of the HSEs in the martian mantle [Harper et al., 1995; Foley et al., 2005] when the internal heating regime was sufficient to cause wholesale convection in the mantle. A late (ca. 4.4 Ga) delivery, on the other hand, ought to result in an uneven (possibly patchy) distribution of the HSEs and other isotopic signatures [Day et al., 2012]. Relatively late addition to an already differentiated Mars might be expected to lead to tungsten isotopic anomalies for impact-affected materials. Such anomalies have been observed in lunar samples and infer the existence of late accretion to the Moon [Kruijer et al., 2015; Touboul et al., 2015]. Our testable prediction from this model is that different components of different ages of different origins (exogenous vs. endogenous) in the oldest martian meteorites, such as NWA 7034 mentioned above, may be expected to preserve strongly contrasting three-oxygen isotope values (expressed in the conventional $\Delta^{17}O$ notation) well outside the contemporary martian range. It is worth noting that Agee et al. [2013] documented the existence of multiple martian oxygen reservoirs based on its isotope heterogeneities in NWA 7034. We predict that further studies with different isotopic systems (e.g. $^{182}W$) in the oldest components of the martian meteorites will yield a similar outcome. Such a result favors a relatively late, colossal, impact event.

An impact of this magnitude would also be expected to eject a substantial amount of material into orbit around Mars, which could then be the source material that eventually formed its satellites [Craddock, 2011; Citron et al., 2015; Rosenblatt & Charnoz, 2012; Rosenblatt et al. 2016]. Smooth-particle hydrodynamics (SPH) impact simulations onto Mars of a body with 2.6% of the mass of Mars yield a circum-martian debris disc of 1-4% of the impactor's mass (i.e. 0.026 to 0.1 martian wt.%) [Citron et al., 2015]. Such a disc is composed of a mixture of materials from both the impactor target, with 20-70% coming from Mars [Rosenblatt et al., 2016; Canup & Salmon, 2016]. The circum-martian disc is then thought to coagulate into a series of satellites, of which Phobos and Deimos are the last two that remain [Rosenblatt & Charnoz, 2012; Rosenblatt et al., 2016]. A recent analysis of the cratering on Phobos indicates it could be older than ca. 4.3 Ga [Schmedemann et al., 2014], lending tentative support to the very early impact-generated formation of Mars' satellites.

The composition, age and origins of the martian satellites remains controversial. Phobos' high porosity appears inconsistent with an origin from a differentiated body, and the satellite's density and spectrum supports an asteroidal origin [Andert et al., 2010; cf. Rosenblatt et al., 2016]. Non-martian material in the disc can be reconciled with the giant collision scenario and the primitive composition of the martian satellites suggested by remote sensing observations of their surfaces [Murchie et al., 2015], although such measurements are only representative of the surface, not of the bulk composition of the bodies.

## 5 Conclusions

After their final assembly, all of the terrestrial planets experienced late accretion from planetesimals left over from planet formation. Some of this late accretion occurred at a time when core separation in the inner planets had been mostly completed. After crust formation was complete, some of the highly siderophile elements that were delivered in late accretion ended up suspended in the mantles of the terrestrial planets rather than in their cores. The amount of late accretion on Mars as traced by HSEs is still uncertain, but best estimates show it was approximately 0.8 wt.%. Through a combination of *N*-body and Monte Carlo experiments we show that if Mars indeed suffered this much late accretion then it must have experienced an impact with a Ceres-sized planetesimal. This impact occurred early, before the onset of global crust formation near 4.43 Ga, and is possibly responsible for the hemispherical dichotomy as well as the origin of its satellites. A cryptic signal of this event may be preserved in the oldest components of the martian meteorite clan in the form of isotope heterogeneities.

**Acknowledgments**

The authors thank James Day and an anonymous reviewer for their feedback that greatly improved this manuscript. RB is grateful for financial support from JSPS KAKENHI (JP16K17662). RB and SJM acknowledge the Collaborative for Research in Origins (CRiO) which is supported by The John Templeton Foundation – FfAME Origins program: The opinions expressed in this publication are those of the authors, and do not necessarily reflect the views of the John Templeton Foundation. SJM is also grateful to the NASA Exobiology Program (NNH14ZDA001N-EXO awarded to Principal Investigator S. Marchi) which helps support our investigations of terrestrial-type planetary bombardments. The source codes for the model used in this study are archived at the Earth Life Science Institute of the Tokyo Institute of Technology. The data, input and output files necessary to reproduce the figures are available from the authors upon request.

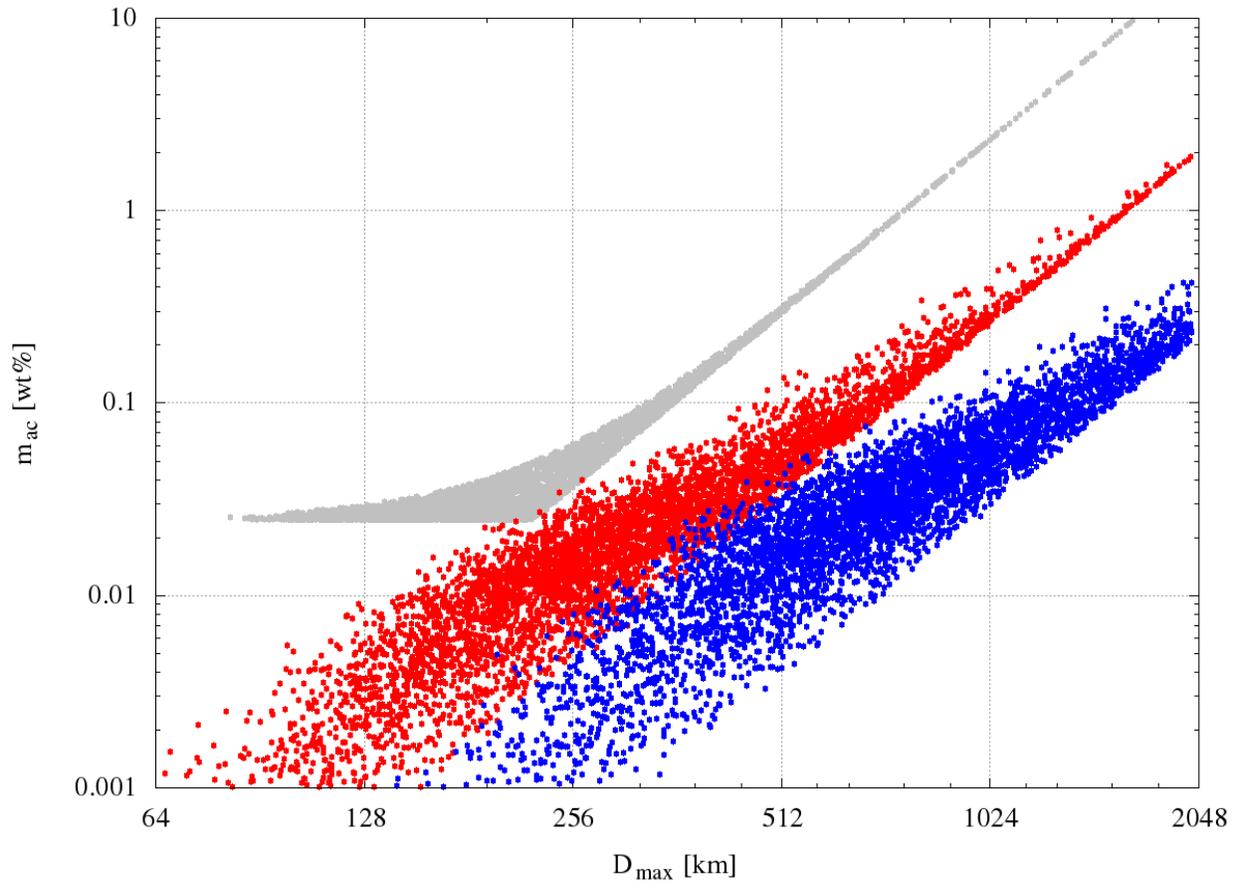

**Figure 1**. The total late accreted mass (in wt.%) versus the diameter of the largest body to strike the Moon (grey), Mars (red) and the Earth (blue). The plateau seen for the Moon near 0.025 wt.% is caused by the requirement that the Moon experienced at least this much late accretion. In this experiment the diameter of the largest body was set to 2000 km. For the Earth to accrete 1 wt.% requires the impact of a lunar-sized body. If Mars experienced all its 0.7 wt.% late accretion after Moon formation it should have been struck by a body with a diameter of 1000 km or larger.

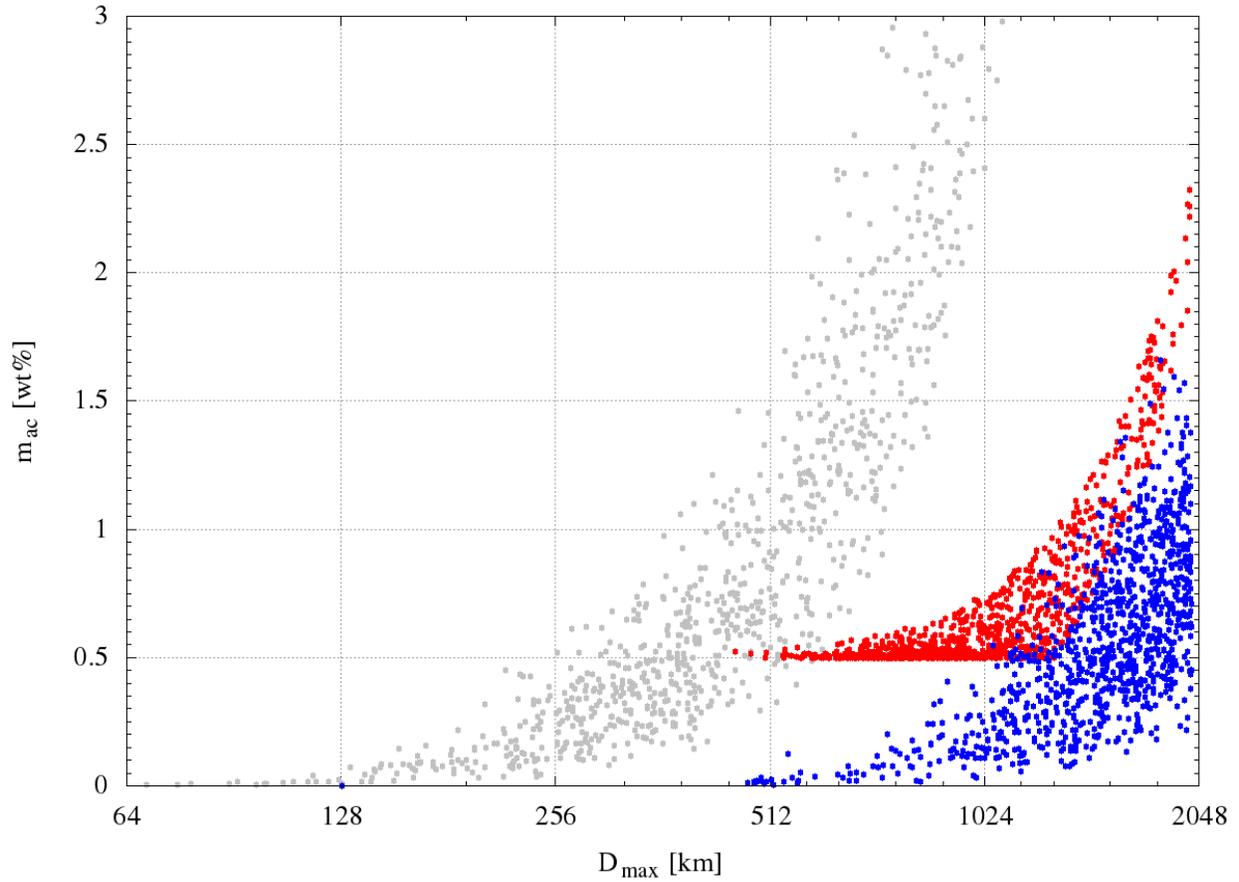

Figure 2. The total late accreted mass (in wt.%) versus the diameter of the largest body to strike the Moon (grey), Mars (red) and the Earth (blue). The plateau seen for Mars at 0.5 wt.% is caused by the requirement that it experienced at least this much late accretion. In this experiment the diameter of the largest body was set to 2000 km. The data indicate that Mars should have been struck by a body with a diameter of 1000 km or larger if it experienced 0.7 wt.% late accretion.

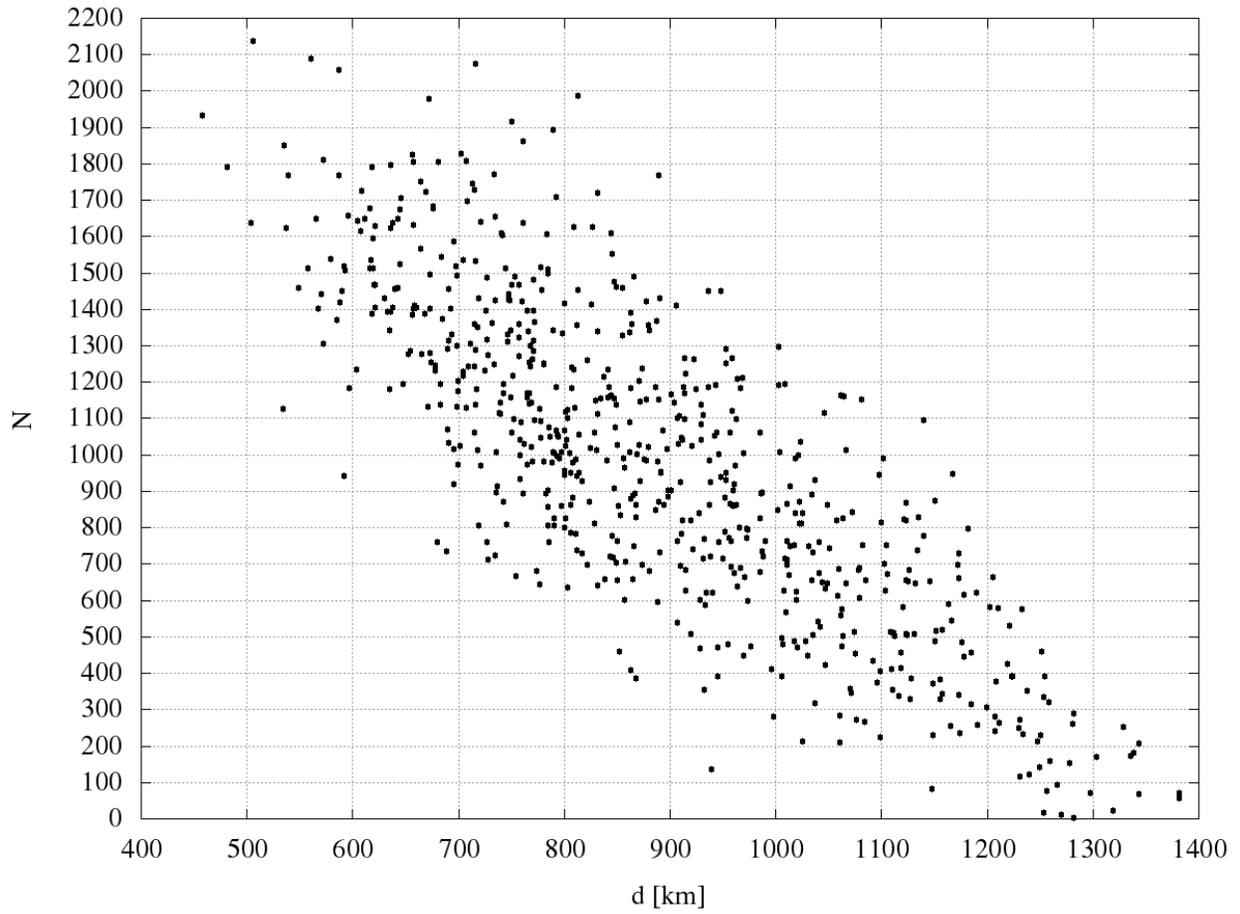

**Figure 3**. The number of expected martian basins versus the diameter of the largest impactor. The current number of basins is only reproduced if the diameter of the largest impactor is 1300 km or higher.